\documentclass[12pt,english,eqsecnum,prd,aps,superscriptaddress,longbibliography,tightenlines,nofootinbib]{revtex4-2}

\usepackage[unicode=true,pdfusetitle,bookmarks=true,bookmarksnumbered=false,bookmarksopen=false,breaklinks=false,pdfborder={0 0 1},backref=false,colorlinks=true]
{hyperref}
\hypersetup{allcolors=blue}
 
\usepackage{amsmath,amssymb,bm}
\usepackage{graphicx}
\usepackage{mathtools}
\usepackage{epstopdf}
\usepackage{amsfonts}
\usepackage{amssymb}
\usepackage{amsbsy}
\usepackage{amsmath}
\usepackage{latexsym}
\usepackage{sansmath}
\usepackage[caption=false]{subfig}
\usepackage{float}
\usepackage[normalem]{ulem}

\usepackage{bm}
\usepackage[dvipsnames]{xcolor}
\usepackage{comment}
\usepackage{csquotes}
\usepackage{physics}
\usepackage{accents}

\def\bea {\begin{eqnarray}}
\def\eea {\end{eqnarray}}
\def\nn {\nonumber}

\begin{document}
 
\title{Dynamical model for black hole to white hole transitions}
\author{Samantha Hergott}
\email{sherrgs@yorku.ca}
\affiliation{Department of Physics and Astronomy, \\York University, Toronto, Ontario M3J 1P3, Canada}
\affiliation{Perimeter Institute for Theoretical,\\ 31 Caroline St N, Waterloo, Ontario N2L 2Y5, Canada}

\author{Viqar Husain}
\email{vhusain@unb.ca}
\affiliation{Department of Mathematics and Statistics, University of New Brunswick,\\
Fredericton, NB, Canada E3B 5A3}

\author{Saeed Rastgoo}
\email{srastgoo@ualberta.ca}
\affiliation{Department of Physics, Theoretical Physics Institute,\\ \& Department of Mathematical and Statistical Sciences,\\ University of Alberta, Edmonton, Alberta T6G 2G1, Canada}

\begin{abstract}
\vskip 0.5cm

We present an asymptotically flat spherically symmetric non-singular metric that describes gravitational collapse and matter bounce with transient black hole and white hole regions. The metric provides a dynamical counterpart to proposed static non-singular black holes, and a phenomenological model for possible black hole to white hole transitions in quantum gravity.  

\end{abstract}

\maketitle

\section{Introduction}

A complete theory of quantum gravity remains elusive. However, there is broad agreement on some of its expected features. One such expectation, perhaps the most important, is singularity resolution and the final state of black holes.  There are three broad approaches to addressing this question: propose phenomenological metrics using general arguments \cite{Hayward:2005gi, Ashtekar:2005cj,Haggard:2014rza,DeLorenzo:2014pta,Barcelo:2014cla,Barcelo:2015noa,Maeda:2021jdc,Carballo-Rubio:2025fnc} and study their properties without recourse to any specific theory;  quantize the Schwarzschild black hole \cite{Kuchar1994,Modesto:2005zm,Boehmer:2007ket,Ashtekar:2023cod,Fragomeno:2024tlh, Gingrich:2024mgk}, and seek static or dynamical solutions of proposed effective equations with matter that contain quantum gravity corrections \cite{Husain:2008tc,Ziprick:2009nd, Kreienbuehl:2010vc, Ziprick:2016ogy, Husain:2021ojz, Husain:2022gwp, Giesel:2022rxi, Giesel:2023hys}. 

Nearly all approaches use mini/midi superspace models using heuristic arguments, the framework of loop quantum gravity (LQG), or Wheeler-deWitt quantization, and all are in spherical symmetry, with or without matter fields. The outcomes of these investigations are varied: non-singular static black holes (BHs); Planck star remnants;  black hole to white hole (WH) transitions; and matter bounce resulting in an outgoing shock wave.  

Since black holes form from collapsing matter, and quantum gravity corrections are expected to provide a matter bounce,  e.g. as seen in models of dust collapse \cite{Husain:2021ojz}, we take the perspective that models with matter are crucial to understanding the fate of black holes in quantum gravity, and that vacuum models are insufficient.  

A natural way to construct effective metrics, whether directly or by solving effective equations, is to ``follow the matter.'' Following this idea, in an earlier work the authors introduced a smooth dynamical metric that describes the collapse and bounce of matter exhibiting formation and evaporation of black hole horizons \cite{Hergott:2022hjm}. The model uses generalized Painlevé-Gullstrand (PG) coordinates and is constructed using a mass function that describes matter inflow and outflow without using junction conditions. The mass flow is governed by a velocity function with parameters that can be adjusted to obtain a black hole lifetime proportional to the cube of the black hole mass $M^3$, in accordance with Hawking evaporation, as well as other powers of $M$. The metric contains a transient trapped region without a spacetime singularity, but does not contain anti-trapped regions. As such, it may be viewed as a dynamical analog of proposed static regular black holes.   

A question of interest is whether it is possible to obtain a dynamical metric in a single coordinate system that exhibits a BH $\rightarrow$ WH transition with a single asymptotic region. There are proposed metrics and conformal diagrams for such metrics for such transitions, but most involve gluing two or more metrics (e.g. \cite{Han:2023wxg}) to obtain the desired features, without focus on matter flows. An exception is a model in 2-dimensional dilaton gravity where the semiclassical Einstein equation with a source given by the trace anomaly gives a solution with BH and WH regions \cite{Barenboim:2024dko}. 

Here we present a smooth, dynamical, non-singular and asymptotically flat metric in PG coordinates that describes a dynamical BH $\rightarrow$ WH transition with a single asymptotic region. The metric contains trapped and anti-trapped regions of finite lifetime, and the spacetime may be interpreted as arising from an inhomogeneous and anisotropic fluid. In the context of the current literature in this area, the metric provides perhaps the simplest example of a dynamical singularity-free spacetime with features that may arise from an effective theory with quantum gravity corrections. We note that in addition to the works cited above, our proposal is in a spirit similar to other conjectured metrics with potentially interesting physical properties such as warp drives \cite{Alcubierre_1994} and wormholes \cite{Thornewormhole}. 

In the next section we review our previous model, point out why it does not have an anti-trapped region, present the extension necessary for describing a BH $\rightarrow $ WH transition, and describe the properties of the resulting metric.  We conclude in Sec. \ref{Sec:conclude} with a discussion of our results in light of other proposals.  


\section{Black hole to White hole transition metric\label{Sec: metric}}

In a previous work \cite{Hergott:2022hjm} we studied the metric  
\begin{equation}
ds^{2}=-\left[1-\left(N^{r}(r,t)\right)^2\right]dt^{2}+2N^{r}(r,t)dtdr+dr^{2}+r^{2}d\Omega^{2},\label{eq:metric-final}
\end{equation}
\begin{equation}
N^{r}(r,t)=\sqrt{\frac{2m(r,t)}{r}},\label{eq:shift}
\end{equation}
with mass function $m(r,t)$ given by 
\begin{equation}
m(r,t)=M_{0}\left[1+\tanh\left(\frac{r-r_{0}-v(r,t)\,(t-t_0)}{\alpha\ l_{0}}\right)\right]^{a}\tanh\left[\left(\frac{r}{l_{0}}\right)\right]^b.\label{eq:mass}
\end{equation}
This has parameters $\alpha, a,b$, a radial point $r_0$ that is the location of the bounce, a fundamental length scale $l_0$, and a ``velocity" function 
\begin{equation}
v(r,t)=\frac{A}{(1+r)^{n}}\tanh\left(\frac{t-t_{0}}{l_{0}}\right),\label{v}
\end{equation}
that determines the radial velocity profile. This function, in turn, includes parameters $A>0$ and $n>0$, which control, among other things, how fast the collapse happens. The parameter $t_0$ is the time at which the velocity reverses from ingoing to outgoing. From the mass function it is evident that the Arnowitt-Deser-Misner mass associated to the metric is 
\begin{equation}
M_{\text{ADM}}\coloneqq\lim_{r\to\infty}m(r,t)= 2^{a}M_{0},
\end{equation}
while the matter density is 
\begin{equation}
\rho(r,t)= \frac{1}{4\pi r^2} \frac{\partial m(r,t)}{\partial r}. \label{eq:density}
\end{equation}
The parameters in the metric are chosen such that the density remains finite as $r \to 0$. It is evident by construction that the mass function $m(r,t)$ is monotonic, increases to $M_{\text{ADM}}$ as $r$ increases, and that the radial location of the rise is time-dependent to model the mass inflow and outflow. The null expansions of the metric (\ref{eq:metric-final}) are 
\begin{equation}
\theta_{\pm}=\frac{2}{r}(-N^{r}\pm1).\label{eq:exp}
\end{equation}
Since $N^r$ is positive by its definition as a function of $m(r,t)$, it is clear that while $\theta_+$ is positive at large $r$ and can change sign at small $r$ indicating trapping, $\theta_- <0$ for all $r$. Therefore, the spacetime does not contain a WH region; further details appear in \cite{Hergott:2022hjm}.

It is apparent from (\ref{eq:exp}) that for a metric to have both BH ($\theta_+<0$ and $\theta_-<0$) and WH ($\theta_+>0$ and $\theta_->0$) regions it is necessary that $N^r$ change sign dynamically near the matter bounce time. One way to arrange such a scenario  is by setting
\begin{equation}
N^{r}(r,t)=-\tanh[\lambda(t-t_{0})]\ \sqrt{\frac{2m(r,t)}{r}},\label{eq:new_shift}
\end{equation}
for the mass function in (\ref{eq:mass}) and a parameter $\lambda>0$. The metric (\ref{eq:metric-final}) with this shift function is the model we consider. As we will see, this modified shift vector suffices to describe matter collapse to form a black hole, a bounce to a white hole region, followed finally by the disappearance of the white hole horizon. The metric remains asymptotically flat and singularity-free as the smooth sign switching function is independent of $r$. We now describe these properties in detail.

\subsection{Transient horizons}

\begin{figure}
\includegraphics[width=\textwidth]{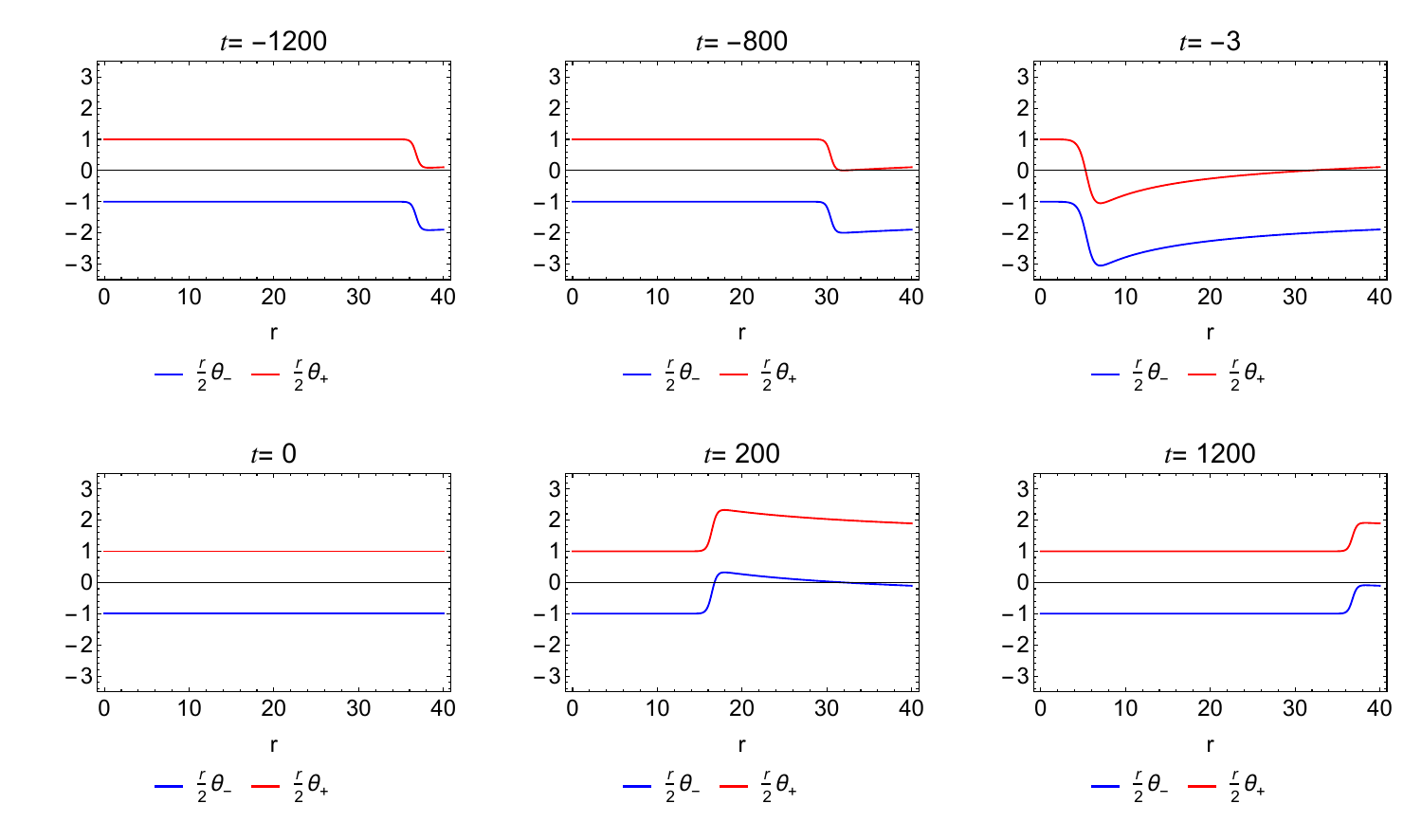} 
\caption{\small{Null expansion scalars $\theta_\pm$ that exhibit a transition from a black hole ($\theta_+<0$ for some range of $r$ and $\theta_-<0$ for all $r$) to a white hole ($\theta_->0$ for some range of $r$ and $\theta_+>0$ for all $r$) at the times indicated: a BH of radius $r=32$ forms at $t=-800$ and disappears at $t=0$; a WH of radius $r=32$ has formed at $t=20$ and disappears at $t=1200$.}}
\label{fig:exp_contour2} 
\end{figure}

The formation and evolution of horizons is demonstrated by computing $\theta_\pm(r,t)$. The roots of these functions give the locations of outer and inner horizons as a function of $t$.  Fig. \ref{fig:exp_contour2} shows typical plots of $r\theta_\pm/2$ for the parameter values $M_{0}=4,\ a=2,\ b=3,\ r_{0}=5,\ t_{0}=0$, with all other parameters set to unity. There are no roots in the first frame $(t=-1200)$ indicating no horizons; the next two frames show roots of $\theta_+$ indicating the formation of a black hole with the outer horizon fixed at $r=2M_{\text{ADM}} = 32$ (for the chosen parameters); the inner horizon moves inward with the matter flow; the fourth frame ($t=0$) is the transition point with no roots and therefore no horizons; the next frame ($t=200$) shows an inner and outer white hole horizon at $r=32$, followed by disappearance of all horizons in the last frame ($t=1200$). 

An important feature in this time sequence is that the transition from a black region ($\theta_\pm<0$) to a white hole region ($\theta_\pm>0$) must be such that the respective regions cannot overlap in spacetime. In the coordinates that we are using, it turns out that there is a spacetime gap near $t=0$ between these regions. The extent of this gap depends on the parameters in the metric, and it may be possible to fine-tune the parameters so that the regions touch at a point. But such a situation is not the general case, at least for this metric.

\begin{figure}
\includegraphics[width=\textwidth]{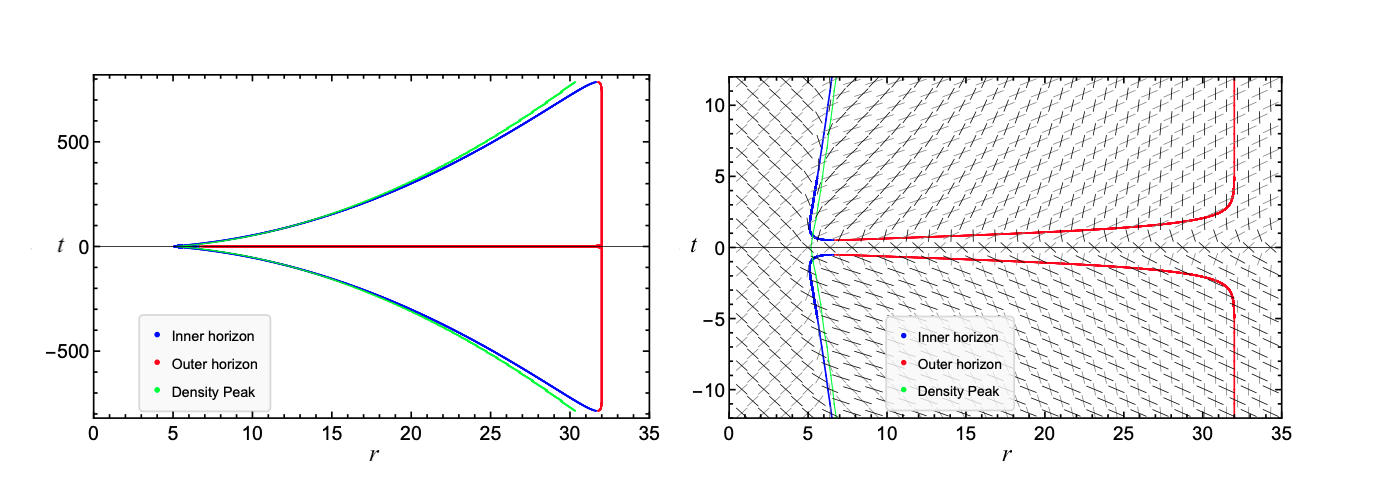} 
\caption{\small{A view of the black hole and white hole regions: the left frame shows the inner and outer horizons and the peak of the density profile; the right frame is a close up showing the separation between the BH and WH regions with a sample of light cones.}}
\label{fig:bh-wh-rt} 
\end{figure}

Another view of this feature is provided in Fig. \ref{fig:bh-wh-rt} which shows the BH and WH regions in the $r-t$ plane: the left frame shows the inner (blue) and outer (red) horizons, and the peak value of the matter density (\ref{eq:density})(green); the right frame is a close-up version that resolves the separation between the BH and WH regions near the matter bounce time $t=0$; also shown are a sampling of light cones. These figures show several notable features: (i) the outer BH horizon turns spacelike, and moves radially inward to meet the inner BH horizon near $t\approx 0.5$;  (ii) the matter density leaves the BH region as the inner and outer BH horizons annihilate near $t=-0.5$, and then enters the WH region while moving in a timelike manner; (iii) during the relatively small time interval between $t=-0.5$ and $t=0.5$, there are no horizons, and one may interpret the metric as that of a transient remnant emerging from the black hole region and then transitioning into a white hole. 

\begin{figure}
\includegraphics[scale=0.35]{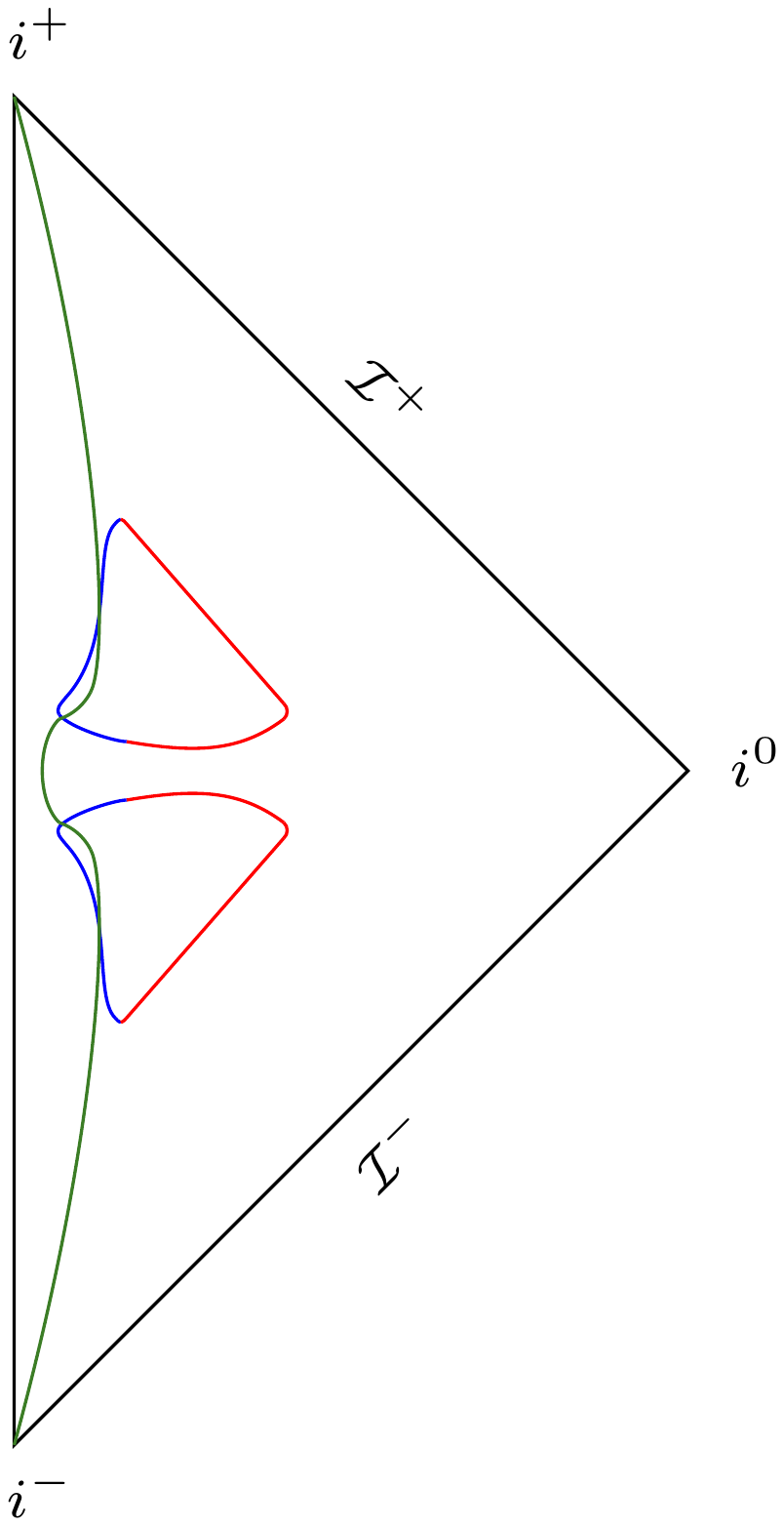} \caption{\small{Schematic conformal diagram of the black hole (lower region) to white hole (upper region) transition. }}
\label{fig:penrose} 
\end{figure}

Fig. \ref{fig:penrose} is a schematic conformal diagram of the spacetime that exhibits these features: the asymptotic structure is identical to that of Minkowski spacetime; the bulk region contains two closed disjoint regions representing the black and white holes; the BH region (lower) is such that the outer horizon (red) is null almost everywhere and the inner one (blue) is initially timelike, and then becomes spacelike as it re-approaches the outer horizon; the WH region (upper) has an outer horizon (red)  that is almost everywhere null and the inner horizon (blue) is initially spacelike and becomes timelike as it rejoins the outer horizon. In this class of metrics there are no parameters such that the outer horizons of the BH and WH regions meet at a point. This would require $\lambda\to \infty$ in \eqref{eq:new_shift} which would convert the switching function into a step function and make the metric non-differentiable at that time. As a consequence the corresponding stress energy tensor would become a derivative of the delta function, an obviously undesirable feature for an effective metric.  (For $\lambda\to -\infty$ the metric becomes  \eqref{eq:shift}, i.e. that of a BH and no WH.)

\subsection{Energy Conditions\label{Sec:energy}}
In effective or modified gravity models with quantum gravity corrections such as singularity avoidance,  the equations of motion, if any, are not those of general relativity. This is the case for the model we study here.  However, it can still be instructive to study the ``effective" stress energy tensor derived assuming Einstein's equation holds, by setting 
\begin{equation}
T_{\mu\nu}=\frac{1}{8\pi }G_{\mu\nu},\label{eq:SET}
\end{equation}
where $G_{\mu\nu}$ is the Einstein tensor calculated using \eqref{eq:metric-final} with \eqref{eq:new_shift}. This is useful because it provides a view of where in the spacetime one or more of the physically motivated  energy conditions are violated. Prominent examples where a similar analysis is carried are the warp drive  \cite{Alcubierre_1994} and wormhole spacetimes \cite{Thornewormhole}.

\begin{figure}
    \centering
    \subfloat[][\small{$\rho-|p_r|$}]{\includegraphics[scale=0.135]{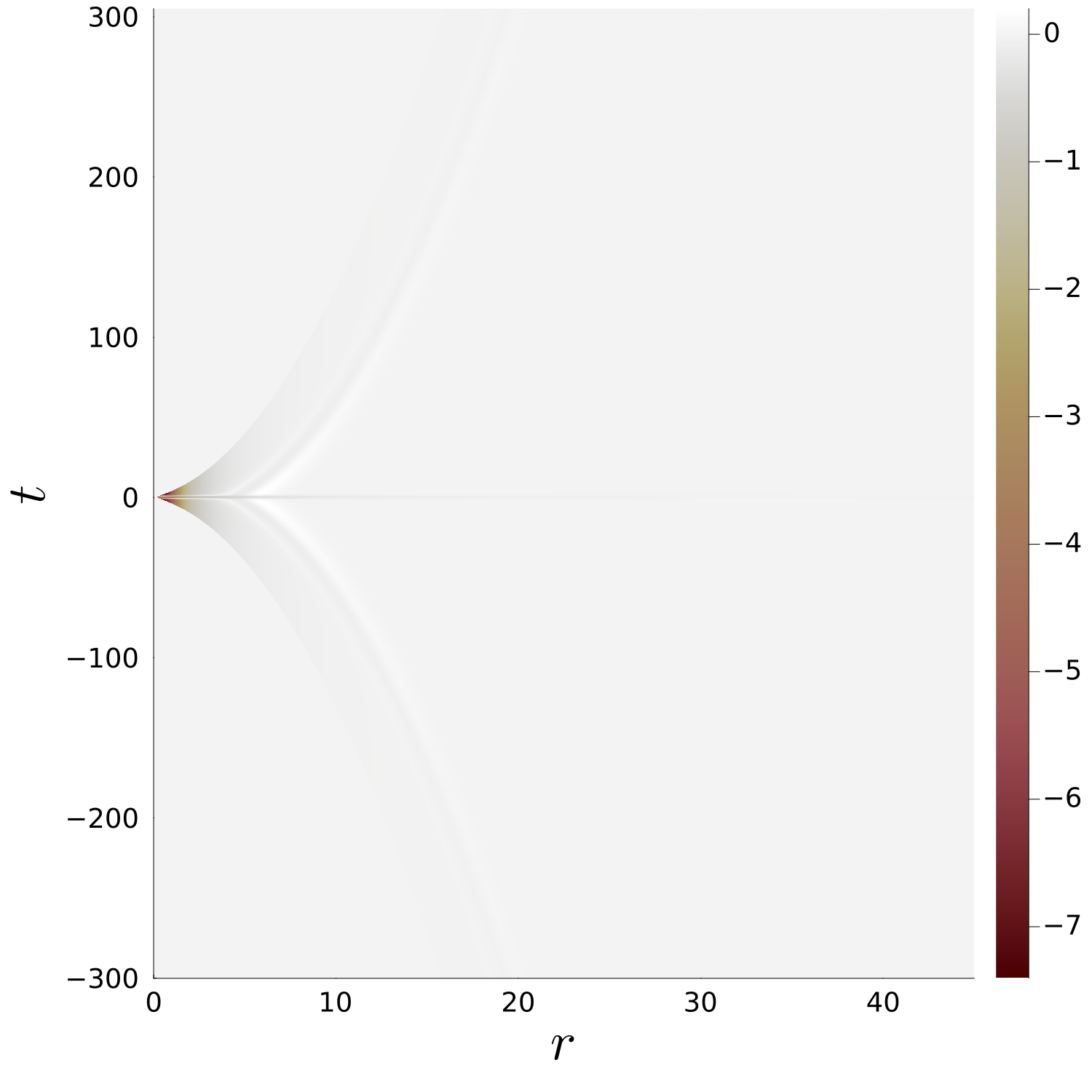} }
    \quad
    \subfloat[][\small{$\rho-|p_\theta|$}]{\includegraphics[scale=0.135]{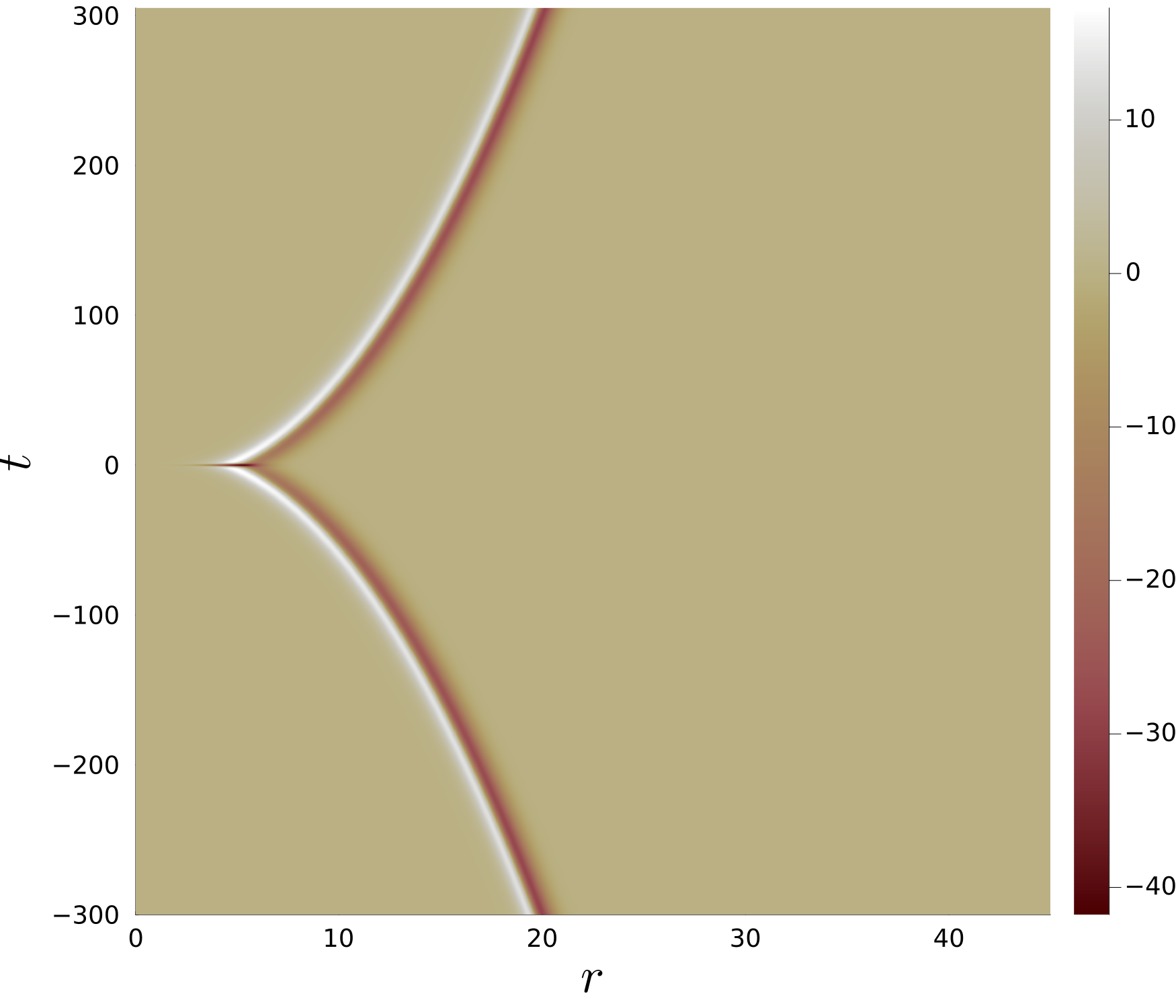}} \\
    \subfloat[][\small{$\rho-|p_r|$}]{\includegraphics[scale=0.135]{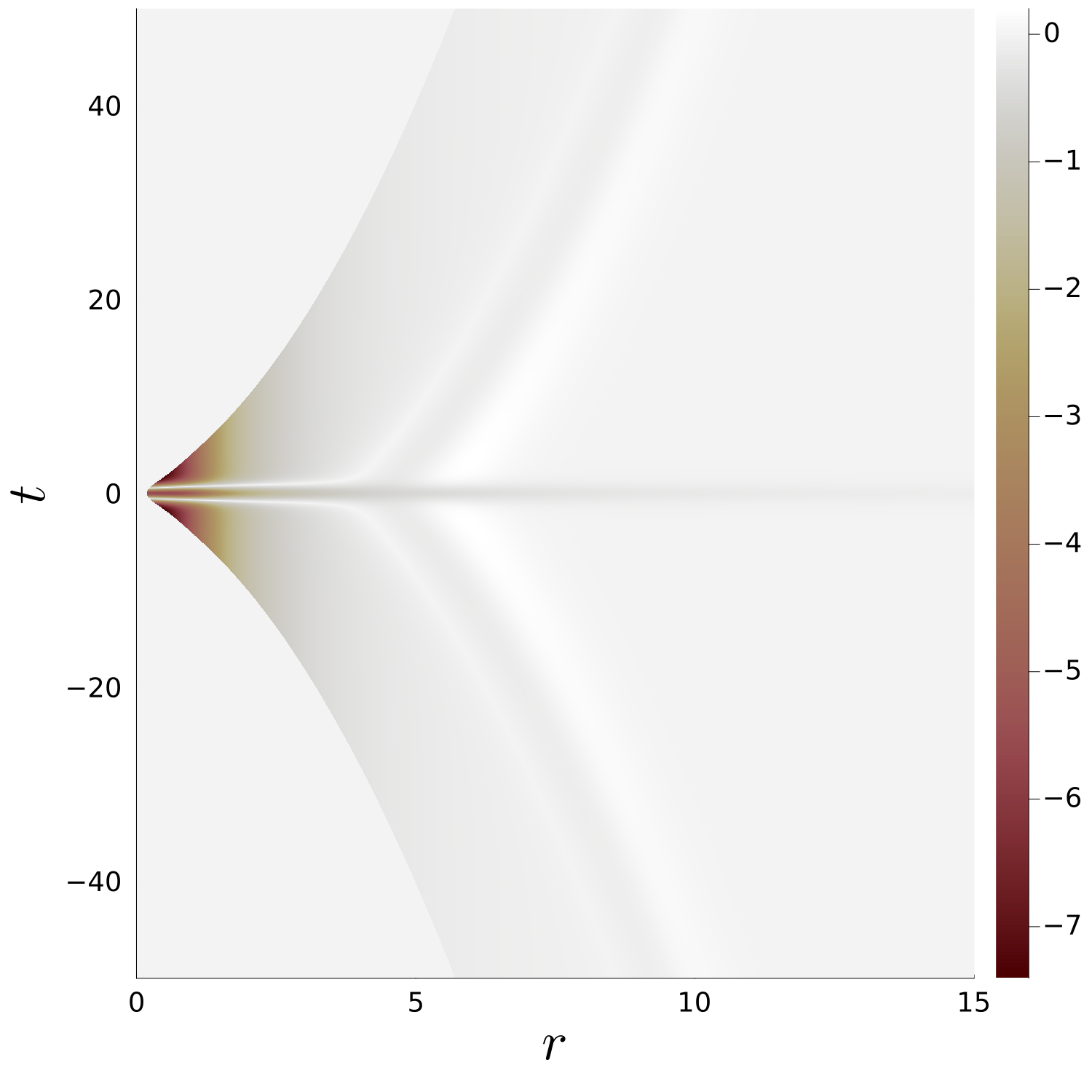}}
   \quad 
  \subfloat[][\small{$\rho-|p_\theta|$}]{\includegraphics[scale=0.135]{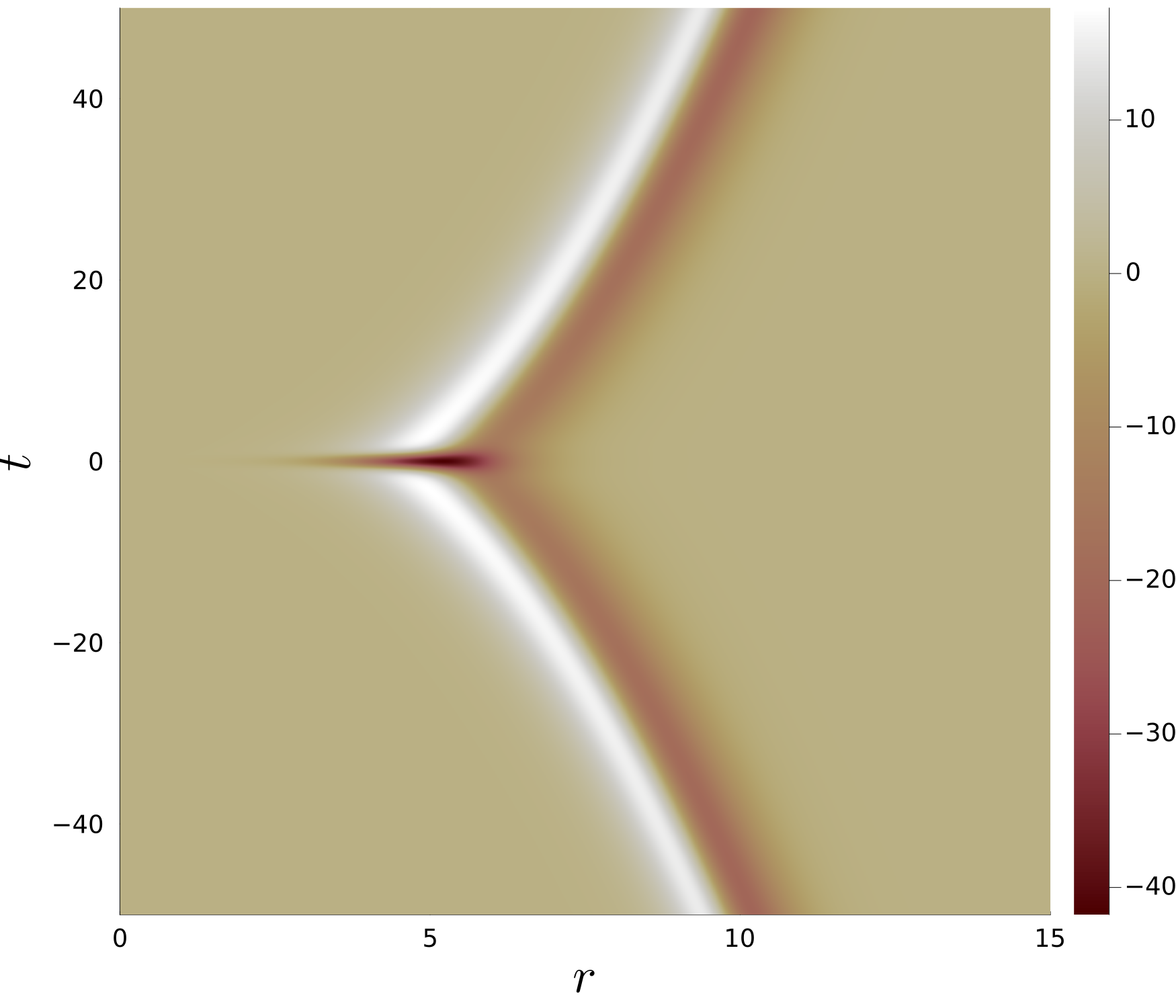}}
    \caption{Radial (a,c) and azimuthal (b,d) components of the dominant energy condition: the second row is a close up of the upper row; the radial direction shows small violations of the DEC near the matter bounce point; the azimuthal direction shows violations at larger radial values.} 
    \label{fig:DEC}
\end{figure}

 For studying the energy conditions, the first step is the derivation of the energy density and principal pressures associated obtained by finding the eigenvalues $T^a_{\ b}w^b = \lambda w^a$ (see, e.g. \cite{Wald:1984rg}). For our metric these are 
 \bea
\rho & =&\frac{1}{8\pi r^{2}}N^{r}\left[2r\left(\partial_{r}N^{r}\right)+N^{r}\right],\\ 
p_r & = &-\frac{1}{8\pi r^{2}}\left[2rN^{r}\partial_{r}N^{r}-2r\partial_{t}N^{r}+\left(N^{r}\right)^{2}\right],\\
p_\theta &=& p_\phi \nn\\
& =& \frac{1}{8\pi r}\left[-r\left(\partial_{r}N^{r}\right)^{2}-N^{r}\left(2\partial_{r}N^{r}+r\partial_{r}^{2}N^{r}\right)+\partial_{t}N^{r}+r\partial_{t}\partial_{r}N^{r}\right].\label{eq:SET_{c}omponents}
\eea
Since the radial and angular pressures are different, an interpretation of these quantities is provided by the stress energy tensor of an anisotropic fluid, 
\begin{equation}
T_{\mu\nu}=\left(\rho+p_{\theta}\right)U^{\mu}U^{\nu}+\left(p_{r}-p_{\theta}\right)W^{\mu}W^{\nu}+p_{\theta}g^{\mu\nu},
\end{equation}
where $U^{\mu}$ and $W^{\mu}$ are respectively normalized timelike and spacelike vector fields with $U^\mu W_\mu=0$.

The dominant energy condition (DEC) reflects the intuition that an observer should see positive energy density and timelike or null energy-momentum flux.  In terms of the quantities above, this gives the inequalities $\rho\geq|p_{r}|$ and $\rho\geq|p_{\theta}|$. Violation of the dominant energy condition is expected because of the matter bounce built into the mass function. Fig. \ref{fig:DEC} shows plots of the quantities $\rho-|p_{r}|$ (left column) and $\rho-|p_{\theta}|$ (right column) for the same metric parameters as for Fig. \ref{fig:bh-wh-rt}; the second row is a close-up of the bounce region near $r=0$ and $t=0$ for each case.  

We see that $\rho-|p_r|$ is approximately zero in the entire spacetime except near the horizon and bounce regions; near the horizon but far from the bounce region, $\rho-|p_r|$ is slightly negative; closer to the bounce, it is significantly negative indicating a stronger violation of the DEC; this continues to larger $r$ on the line $t=0$ as seen in Fig. \ref{fig:DEC}(c), and arises due to the inward motion of the outer BH horizon as it meets the inner horizon; the same observation holds for the mirror reflected WH region. Since the inner horizons track the matter flow,  the latter is spacelike where the inner horizon is spacelike. This explains DEC violation in these regions.  

 The other DEC function $\rho-|p_\theta|$ (second column of Fig. \ref{fig:DEC} is also approximately zero in the entire spacetime except again near the horizons and near the bounce; $\rho-|p_\theta|$ is negative to the right of the horizons (inside the black hole and white hole) and positive to their left (outside but closer to $r=0$ than the inner horizons); near the bounce region, $\rho-|p_\theta|$ takes its most negative value; on the line $t=0$ the only violation of DEC occurs near the bounce region.  (For comparison, in models where the BH $\rightarrow$  WH transition is obtained using junction conditions, energy conditions are violated close to the junction Ref~\cite{Feng:2023pfq}.)

\section{Summary and discussion\label{Sec:conclude}}

We proposed and studied a smooth dynamical spherically symmetric non-singular asymptotically flat metric that describes a BH $\rightarrow$ WH transition. The metric may be viewed as a generalization of \cite{Hergott:2022hjm} where there is matter bounce and black hole evaporation, but without white hole formation. In comparison to other works such as \cite{Han:2023wxg}, the metric we give is constructed by ``following the matter" using a ``designer" mass function in PG coordinates, rather than by gluing together several metrics to achieve the desired outcome. Although the metric captures several desirable features that might emerge from a theory of quantum gravity, it remains to be seen whether it can arise as a solution in an effective theory with quantum gravity corrections. This requires deriving effective equations for gravitational collapse in spherical symmetry. As noted above, such equations have been derived for pressureless dust in LQG, but their solutions do not predict a BH to WH transition. The question is whether collapse of matter with pressure, such as a perfect fluid or scalar field gives rise to WH formation after a bounce. In this respect, the metric we have presented can be viewed as a phenomenological model that describes a plausible dynamics that could arise from an actual effective theory. As noted above, this is similar in spirit to several other mostly static metrics for wormholes and dynamical metrics using junction conditions.

As a general observation concerning matter bounce with or without WH formation, the fact that inner horizons become spacelike with the matter flow that defines the geometry means that violation of DEC is inevitable. The only way a matter bounce (and accompanying spacelike flux) can be avoided is if the bounce remains behind a permanent outer (event) horizon, a situation that may be called ``superluminal censorship'' --- a feature that arises in certain weak solutions for classical dust collapse \cite{Husain:2025wrh}, but not for weak solutions of effective models \cite{Husain:2021ojz,Liu:2025fil} where matter emerges as a shock wave as the black hole ends. These no-WH cases have a conformal structure similar to that for the smooth metrics studied in \cite{Hergott:2022hjm}, but it remains to be seen whether the type of BH $\rightarrow$ WH spacetime we give here can arise in an effective theory.    

There are a few specific features of the geometry that are noteworthy: (i) the BH and WH horizons do not intersect at a point, unlike what is proposed in models that use junctions conditions; (ii) the matter and inner horizons move together and are timelike in some regions and spacelike in others; (iii) matter is superluminal within the BH and WH, but becomes timelike in the gap region between them (Fig. \ref{fig:bh-wh-rt}); (iv) the DEC is violated only in the spacetime region where the bounce occurs, as expected in any bounce scenario, but the energy density $\rho$ remains positive everywhere; in particular, the DEC is not violated at sufficiently large radial coordinate.

On the first of these observations, it is useful to contrast with the ``conventional" conjectured picture of a transition where the BH and WH horizons intersect at a point.  In contrast, in metric we present, there is a short-lived gap between the BH and WH regions. We think this is a required feature of a dynamical transition because the condition for a single point of contact $(t_c,r_c)$ between the BH ($\theta_-<0, \ \theta_+<0$)  and WH ($\theta_->0, \ \theta_+>0$) regions would require $\theta_+(t_c,r_c)=\theta_-(t_c,r_c)=0$, which from (\ref{eq:exp}) leads to an obvious contradiction. We note further that the qualitative features of the BH to WH transition are unaffected and robust if other choices of smooth sign switching of the shift function are used.

It is possible that there are astrophysical implications of the metrics we proposed. A first step in this direction is to study geodesics to see if there are possible observational signatures of the transition, especially of the short-lived ``remnant" between the BH and WH regions. That remnants might provide a contribution to dark matter, along with primordial black holes, has been discussed in the literature \cite{Chen:2002tu,nwgd-g3zl}.

 \acknowledgments{
This work was supported in part by  Natural Science and Engineering Research Council of Canada grants to V. H. and S.R.;   S. R. acknowledges the support of the Natural Science and Engineering Research Council of Canada, funding reference No. RGPIN-2021-03644 and No. DGECR-2021-00302.}

\appendix
\bibliographystyle{apsrev4-2}
\bibliography{mainbib}

\end{document}